\documentclass[twocolumn,preprintnumbers,amsmath,amssymb,aps,pra,showpacs,floatfix,superscriptaddress]{revtex4-1}
\usepackage{natbib}
\usepackage{graphicx}
\usepackage{epsfig}
\usepackage{dcolumn}
\usepackage{bm}
\usepackage{soul}
\usepackage{sidecap}
\usepackage[usenames,dvipsnames]{color}
\usepackage{setspace}
\usepackage[colorlinks=true,linkcolor=blue,urlcolor=blue,citecolor=blue]{hyperref}
\usepackage[usenames,dvipsnames,svgnames,table]{xcolor}
\usepackage{wasysym}
\usepackage{multirow}
\usepackage{ulem}

\def\STO{SrTi$_{1-x}$Nb$_{x}$O$_{3}~$}
\def\hc1{$H_{c1}$}
\def\1020{$10^{20}$ cm$^{-3}$}

\begin{document}
\title{Superfluid density and carrier concentration across a superconducting dome: the case of strontium titanate}
\author{Cl\'ement Collignon $^{1,2}$,   Beno\^{\i}t Fauqu\'e$^{1,3}$, Antonella Cavanna$^{4}$ ,  Ulf Gennser$^{4}$,  Dominique Mailly$^{4}$ and Kamran Behnia$^{1}$\email{kamran.behnia@espci.fr}}
\affiliation{(1) Laboratoire Physique et Etude de Mat\'{e}riaux (UMR 8213-CNRS), ESPCI, UPMC, PSL,  F-75005, Paris, France\\
(2) D\'epartement de physique and RQMP, Universit\'e de Sherbrooke, Sherbrooke, Qu\'ebec J1K 2R1, Canada\\
(3) JEIP,  (USR 3573 CNRS), Coll\`{e}ge de France, PSL Research University, 11, place Marcelin Berthelot, 75231 Paris Cedex 05, France\\
(4) Centre de Nanosciences et de Nanotechnologies, CNRS, Univ. Paris-Sud, Univ. Paris-Saclay, 91120 Palaiseau, France 
}
\date{October 3, 2017}

\begin{abstract}

We present a study of the lower critical field, \hc1, of \STO as a function of carrier concentration with the aim of quantifying the superfluid density. At low carrier concentration (i.e. the underdoped side), superfluid density and the carrier concentration in the normal state are equal within experimental margin. A significant deviation between the two numbers starts at optimal doping and gradually increases with doping. The inverse of the penetration depth and the critical temperature follow parallel evolutions as in the case of cuprate superconductors. In the overdoped regime, the zero-temperature superfluid density becomes much lower than the normal-state carrier density before vanishing all together. We show that the density mismatch and the clean-to-dirty crossover are concomitant.  Our results imply that the discrepancy between normal and superconducting densities is expected whenever the superconducting gap becomes small enough to put the system in the dirty limit. A quantitative test of the dirty BCS theory is not straightforward, due to he multiplicity of the bands in superconducting strontium titanate.
\end{abstract}
\maketitle

\section{Introduction}

In many superconductors with an insulating parent, the critical temperature is a non-monotonic function of carrier concentration. The very existence of such a superconducting dome  raises a fundamental question. How does the superfluid density, n$_S$, evolve in such a context? Does it remain equal to the concentration of electrons in the normal state? Or does it follow the non-monotonic variation of the critical temperature?  In the case of high-$T_c$ cuprates, the doping dependence of the superfluid stiffness\cite{uemura1989,homes2004,homes2005} has remained the subject of an intense debate, focused on the link between critical temperature and superfluid stiffness\cite{bernhard1995,locquet1996,lemberger2011,deepwell2013,bozovic2016}.   A recent subject of debate has been the correlation between superfluid density and critical temperature in overdoped cuprates. It has been interpreted as incompatible with the standard Bardeen–Cooper–Schrieffer (BCS) description\cite{bozovic2016} or in good agreement with the dirty BCS theory\cite{lee-hone}. However, a comparison between the magnitude of n$_S$  with the normal-state carrier concentration, n$_H$ was absent in this debate. To the best of our knowledge, and in spite of abundant experimental data, such a textbook\cite{tinkham1975} link has never been verified in any superconductor.

Among doped semiconductors with a superconducting ground state\cite{bustarret2015}, SrTiO$_3$\cite{schooley1964,schooley1965,koonce1967,binnig1980,lin2014} distinguishes itself. With only 10$^{-5}$ electron per formula unit (f.u.), it becomes a superconductor and when carrier density exceeds 0.02/f.u., it ceases to be so. The existence of this superconducting dome raises many questions: how does superconductivity persist in the dilute limit in spite of a hierarchy inversion between Fermi and Debye temperatures? Why does it disappear on the overdoped side despite the steady increase in the electronic density of states? Do plasmons play a role in binding Cooper pairs in the extreme dilute limit\cite{takada1980,ruhman2016}? The vicinity to a ferroelectric instability has motivated theoretical scenarios invoking ferroelectric quantum criticality\cite{rowley2014,edge2015}, which have found support in a number of recent experiments\cite{stucky2016,rischau2016}.

On the other hand, the case of SrTiO$_3$ offers a unique opportunity to explore the behavior of superfluid density when the critical temperature is not a monotonic function of carrier density. The superconducting instability occurs in a well-documented Fermi surface in which carrier concentration is known with a reliable accuracy\cite{lin2014} and can be tuned across orders of magnitude. Here, we present an extensive study of the lower critical field in  \STO across the superconducting dome with a focus on the relative magnitude of superfluid and normal-carrier density. In the underdoped regime, we find that the superfluid density extracted from the magnitude of the lower critical field is in  agreement with the carrier concentration in the normal state. Deep in the overdoped regime, a mismatch between the extracted superfluid density and the concentration of normal electrons is detectable and steadily increases with doping. We show that this mismatch is concomitant with the passage from clean to dirty limit. However, we fail to achieve a quantitative account in the dirty limit in a single-band picture. This is most probably because the multiplicity of the electronic bands significantly affects the n$_S$/n$_H$ ratio. The results have implications beyond the case of strontium titanate. Comparing SrTiO$_3$ with a dense s-wave superconductor, namely niobium, we find that when a superconductor is clean, the magnitude of the penetration depth correlates with its carrier density. It is also instructive to compare optimally-doped YBCO with these two systems. Its reported penetration depth in the $ab$ plane happens to be where it is expected according to its carrier density, the effective mass of its carriers and the BCS theory.

\section{Experimental}

\begin{figure}
\centering
\includegraphics[scale=0.4]{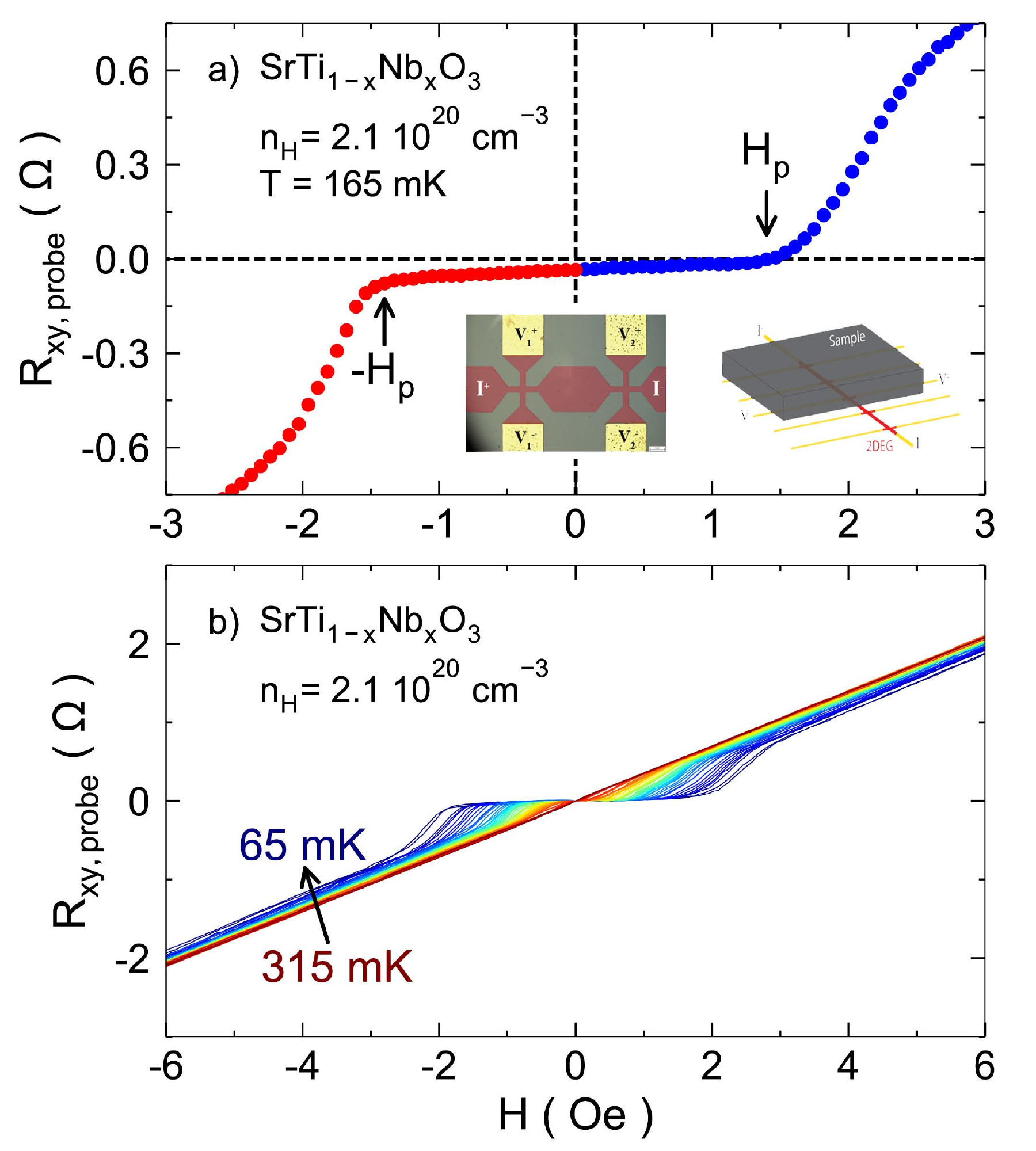}
\caption{ \textbf{a}: Typical field dependence of the Hall resistivity of a  microprobe with a superconducting sample above. The measured signal, directly proportional to the local magnetic field, suddenly rises when the first vortex penetrates the sample. The sharpness  indicates the absence of  surface-barrier effects or gradual vortex leak from sample corners\cite{liang1994}. The inset shows, in false colors, a picture of the array of Hall microprobes used for this experiment and a scheme of how a sample is mounted on the array of probes. \textbf{b}: Raw data for the $n_H = 2.1$ \1020 sample at different temperatures. }
\label{fig1}
\end{figure}

We measured the lower critical field, \hc1, with Hall probes realized using a high-mobility AlGaAs/GaAs heterostructure with a two-dimensional electron gas (2DEG) 160 nm below the surface.
The 2DEG has a mobility of 320.000 $\mathrm{cm^2/V s}$ and a carrier density of $2\,10^{11} \mathrm{cm^{−2}}$ at liquid helium temperature.
The devices were fabricated using electron beam lithography and 250V argon ions to define the mesa.
As shown in the inset of Fig.\ref{fig1}.a, each probe is a 5$\times$5 $\mathrm{\mu}$m$^2$ square, 100 $\mathrm{\mu}$m spaced from its immediate neighbor.

Such a device can monitor local magnetization at micron scale and was used before to study vortex avalanches in superconducting niobium\cite{behnia2000}. Similar Hall microprobes have been used before to measure \hc1 in heavy-fermion\cite{okazaki2010} and in iron-based\cite{putzke2014} superconductors.

The Hall resistance of a 2DEG probe yields the local magnetic field, $B$ :
\begin{equation}
R_{xy,probe} = B^2/ne
\end{equation}

Here $n$ is the carrier density of the electron gas. In our case with $1/ne \sim 0.3 \mathrm{\Omega} / \mathrm{G}$, we could easily resolve very small variations in magnetic field.
We checked that the $1/ne$ coefficient does not change with temperature below 1 K. To measure $H_{c1}$, the sample is laid on the array as depicted in the inset of Fig.\ref{fig1}a.
At low field, the sample is in the Meissner state and the probes below do not feel any magnetic field, which is screened by the sample. When the first vortex penetrates the sample, the microprobe detects a rise in the measured magnetic field. The field value at which it occurs gives the penetration field $H_p$. The data for one sample at one temperature, presented in Fig.\ref{fig1}a, illustrates how clearly one can detect $H_p$. Each isothermal curve was obtained by two sets of sweeps from zero to positive and  from zero to negative fields. Between the two sweeps the sample was heated up to a temperature significantly larger than the critical temperature and back to the measuring temperature.
The intrinsic lower critical field, \hc1 is proportional $H_p$. In the case of a platelet, the geometrical factor is:

\begin{equation}
H_{c1} = \frac{1}{\tanh(\sqrt{0.36t/w})}H_p
\end{equation}
Where $t$ and $w$ are respectively the thickness and the width of the slab \cite{brandt1999}. This geometrical factor is given for each sample in Table \ref{characterization}.

We monitored the Hall resistance of several probes in the proximity of the sample edges. Our data is based on a probe located below the sample and about 150 $\mathrm{\mu m}$ distant from the edge. We chose this option to avoid two phenomena, which can disrupt an accurate determination of $H_{c1}$. Vortices can penetrate by the corner of the sample below $H_p$\cite{liang1994}. Therefore, the sample corners are to be avoided. On the other hand, vortex pinning may generate a non-uniform distribution of magnetic field\cite{bean1964_2}. Like a previous study\cite{okazaki2009}, the probe was located close to the edge in order to avoid this. The sharpness of the increase in $R_{xy,probe}$ at $H_p$ in our data indicates that these phenomena do not contaminate our measurements.

A perfect surface may forbid the flux lines to penetrate the sample and thus artificially increase the penetration field\cite{bean1964}. Experimentally, this surface barrier effect will be manifested as a slow increase above $H_p$ and then a sharp rise when the barrier is overcome\cite{liang1994}. Such effect is absent from our data (see Fig.\ref{fig1}a). Most probably, this is because the surface roughness of our samples is of the same order as the penetration depth\cite{burchalov1992}.

Another experimental challenge is the presence of residual and Earth magnetic field. Their magnitude are not negligible compared to the lower critical field measured here. We compensated them by applying a small magnetic field.
As seen in Fig.\ref{fig1}, the magnitude of $H_p$ was identical for the two field polarities. The small asymmetry for the most overdoped sample quantifies the limits of our compensation method.

Still using the Hall probes, we were also able to quantify the upper critical field of our samples.
At a given fixed field, $H_{app}$, when the temperature is decreased below $T_c (H_{app})$ the sample start to expel the magnetic field.
Thus, for a probe under the sample, the signal will be constant at $T > T_c (H_{app})$ and then start to drop in the superconducting phase.
By repeating this measurement at several fixed fields, we can finally extract the field dependency of $T_c$, or the other way around, the temperature dependency of $H_{c2}$. In practice, for applied field larger than 10 Oe, the expelled field is two or three orders smaller than the applied one.
Hence, the drop we want to observe in $R_{xy,probe}$ is below the resolution of the range we have to use to not get a saturation of the signal.
We consequently measured the difference between the signals of two probes, one under the sample and another outside of it, in order to cancel the large and uninteresting signal from the applied field. This method give us directly a signal proportional to the magnetic susceptibility.

We measured six \STO samples (Source: CrysTec GMbH) with labeled niobium concentrations of 0.2, 0.8, 1, 1.4, 2 and 2 \%. Two slabs were cut from each sample: one to measure resistivity and Hall effect and one for the magnetometry measurement of approximate size $1\times1\times0.5$ mm$^3$ (except for one sample with a thickness of 1 mm).
We performed resistivity temperature sweeps from $T=300$ to 2 K at zero field and Hall effect field sweeps from $H=0$ to $\pm12$ T at $T=2$ K via a Quantum Design Physical Property Measurement System (PPMS).
The extracted $\rho_0$ and $n_H$ values are listed in Table \ref{characterization}. Similar samples were studied previously using  multiple experimental techniques, such as electrical resistivity\cite{lin2015Science}, thermal conductivity\cite{linprb2014}, thermoelectric response\cite{lin2013} and specific heat\cite{linprb2014}. Thanks to their high mobility, quantum oscillations can be detected in a moderate magnetic field\cite{lin2013,lin2014}. Their Nb content was checked by secondary ion beam mass spectroscopy (SIMS)\cite{linprb2014}.

\begin{table}[h!]
\begin{center}
\begin{tabular}{|c|ccc|}
\hline
\multirow{2}{*}{x} & $n_H$ & $\rho_0$(2K) & \multirow{2}{*}{$\frac{H_{c1}}{H_p}$} \\
 & ~(10$^{20}$cm$^{-3}$)~ & ~($\mu\Omega$.cm)~ & \\
\hline
0.002&0.41&49&2.32\\
0.08&1.9&53&2.64\\
1&2.1&71&2.57\\
1.4&2.6&109&2.37\\
2&3.2&56&1.78\\
2&3.5&45&2.28\\
\hline
\end{tabular}
\end{center}
\caption{Sample characterization: Hall number, low-temperature resistivity and the geometrical coefficient extracted from width and thickness in order to extract the lower critical field using Eq.2.}
\label{characterization}
\end{table}

\section{Results}
\begin{figure*}
\centering
\includegraphics[scale=0.45]{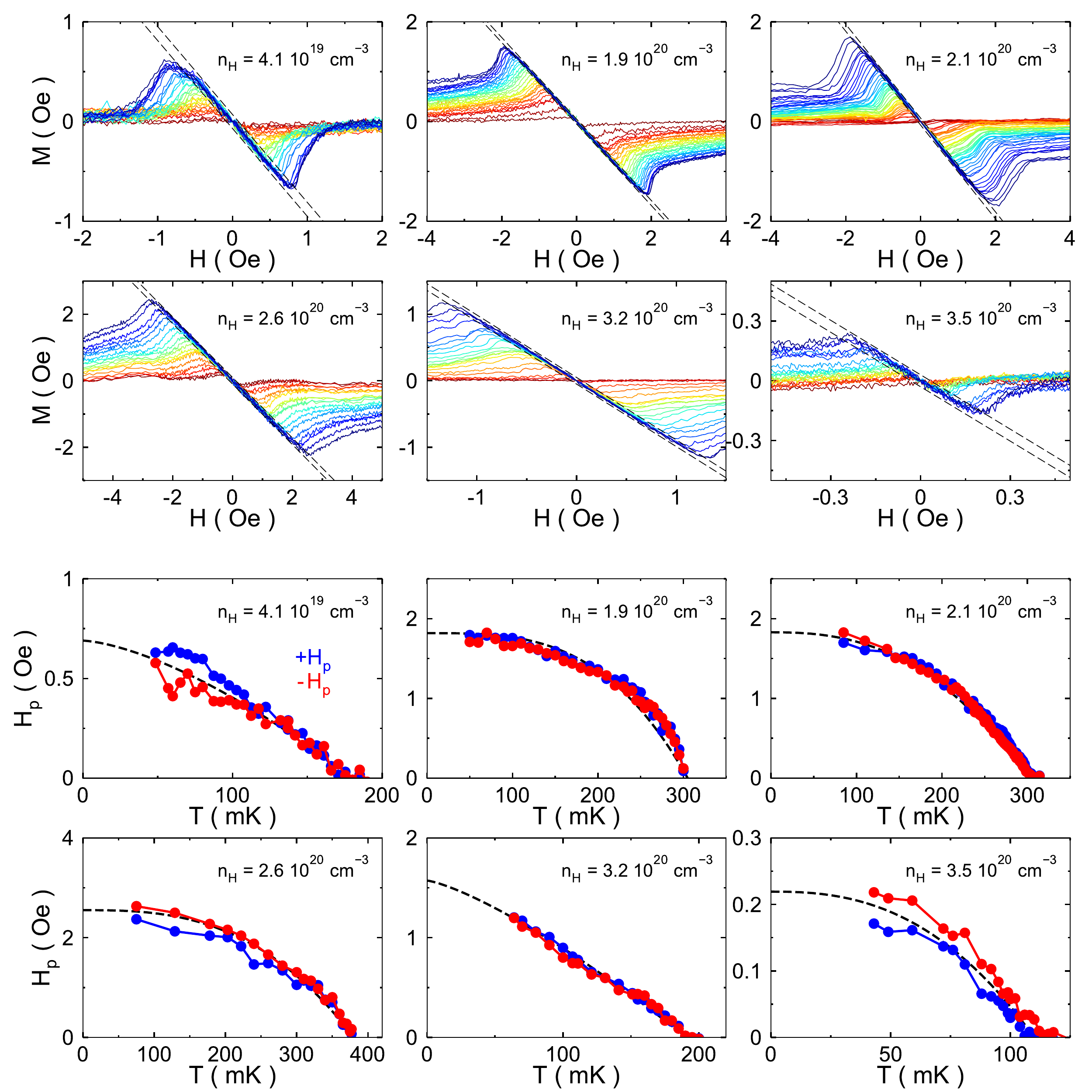}
\caption{Upper panels show magnetization $M$ as a function of magnetic field at different temperatures for the six samples. Lower panels show the extracted penetration fields $H_p$ for opposite orientations of magnetic field by taking the crossing point of M(H) at a given temperature with the dashed lines shown in the panels. Black dash lines are \textcolor{green}{empirical} fits to $H_{p}(T) = H_{p}(0)\,(1-(T/T_c)^{\alpha})$.}
\label{fig2N}
\end{figure*}
\begin{figure}
\centering
\includegraphics[scale=0.4]{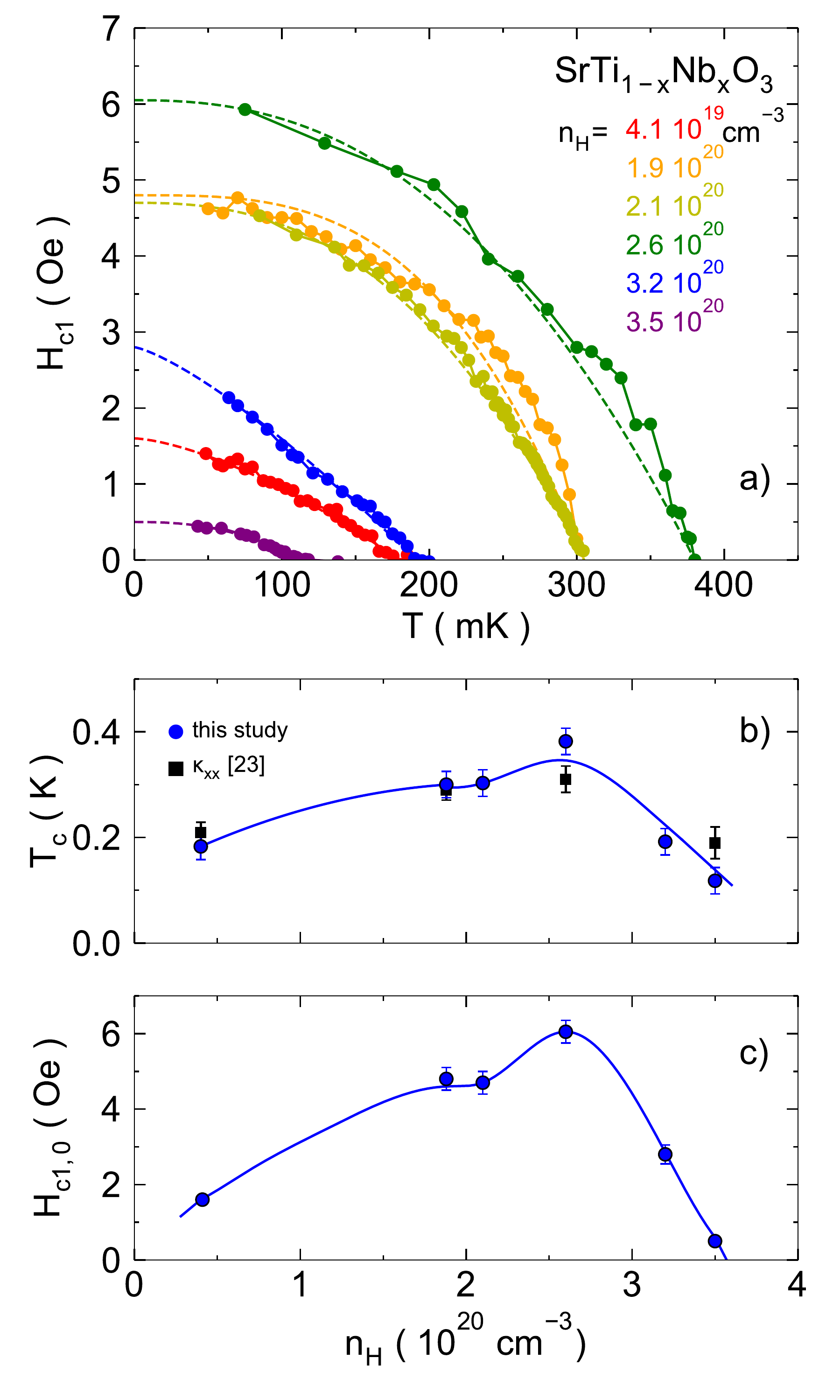}
\caption{\textbf{a}: Lower critical field vs temperature for all six samples deduced from $H_p$ and geometrical correction of the demagnetization field. \textbf{b}: Doping dependence of the critical temperature determined by Hall probe magnetometry (blue disks) and thermal conductivity (black squares)\cite{linprb2014}, the blue line is a guide to the eye.
\textbf{c}: Doping dependence of \hc1, the blue line is a guide to the eye.}
\label{fig2}
\end{figure}
We measured the lower critical field $H_{c1}$ of all six samples and the upper critical field $H_{c2}$ of three of them. These two sets of data allowed us to extract the Ginzburg parameter, $\kappa$. The penetration depth, $\lambda$ can be computed using $\kappa$ and $H_{c1}$. The penetration depth combined with effective mass yields superfluid density, which is the quantity we wish to put under scrutiny.

\subsection{Lower critical field $H_{c1}$}

For each sample, we performed measurements at different temperatures extending from $T_{base} \sim 30 - 50$ mK up to $T_c$ and above, and extracted $H_p(T)$. The raw data for $n_H = 2.1 $ \1020 is shown in Fig.\ref{fig1}b. The measured local field, $B$, can be used to extract magnetization through $ M = B/\mu_0 - H$. In all six samples,  magnetization is proportional to -$H$ at low field, sharply drops at $H_p$ and decreases smoothly afterwards (See Fig.\ref{fig2N}). Superconducting slabs with a small Ginzburg parameter typically behave in this way\cite{brandt1995}. For each temperature, $H_p$ is taken to be the magnetic field at which M deviates significantly from -H (See Fig.\ref{fig2N}). The temperature dependence of $\pm H_p$ for opposite orientations of magnetic field in the six samples is shown in the same figure. The small difference between +$H_p$ and -$H_p$ indicates that residual (including Earth) magnetic field has been compensated in a satisfactory manner.

The temperature dependence of \hc1 for all six samples is exposed in Fig.\ref{fig2}. We note that the temperature dependence of \hc1  is somewhat different among different samples. The  observed variety in the temperature dependence of lower critical field in different samples may be a consequence of multigap superconductivity, which is known to produce additional structure in the temperature dependence of \hc1\cite{askerzade2002}. One may expect that the curvature of \hc1 is set by the way disorder tunes the contribution of different bands. This variety in the temperature dependence has little effect on  the extraction of a reliable H$_{c1}$(0), which is the purpose of the present study. We note an agreement between our data and what was reported in an early study on the lower critical field of two \STO samples near optimal doping\cite{ambler1966}. As one can see in Fig.\ref{fig2}b, the doping dependence of \hc1 and $T_c$ are similar to each other, both presenting a dome-like structure.

\begin{figure}
\centering
\includegraphics[scale=0.45]{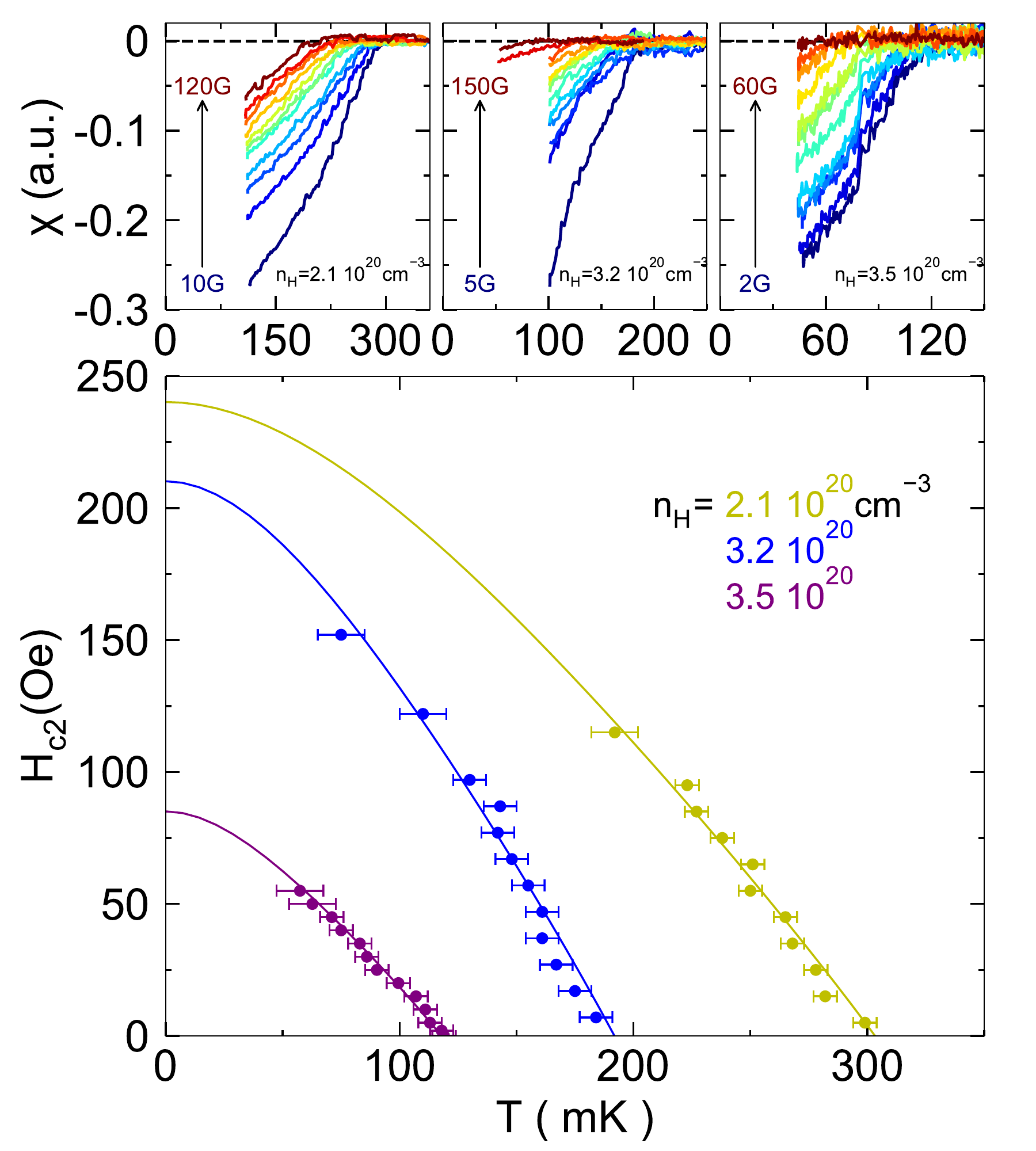}
\caption{Upper panels show the temperature dependence of the difference in signal between one probe under the sample and one probe far from it at various fixed fields and for three dopings.
The difference is proportional to the magnetic susceptibility $\xi$. The lower panel shows the temperature dependence of the upper critical fields $H_{c2}$ extracted from the upper panels. The lines are WHH fits \citep{WHH} to the data points.}
\label{figHc2}
\end{figure}

\subsection{Upper critical field $H_{c2}$}

We managed to measure the upper critical field $H_{c2}$ for three of our samples with carrier densities $n_H = 2.1, 3.2$, and 3.5 \1020 , with the methods explained Part \ref{characterization}.
The signal difference between one probe under the sample and one far from, proportional to the magnetic susceptibility $\xi$, is flat at high temperature and start to drop at $T_c$ as seen in the upper panels of figure \ref{figHc2}.
The field dependence of $T_c$ allows us to plot the temperature dependence of $H_{c2}$ depicted in the lower panel of figure \ref{figHc2}.
We then proceeded to fit the data points with the Werthamer, Helfand and Hohenberg (WHH) function \citep{WHH} in order to extract the value in the $T=0$ limit.
Those fits give the values reported in Table \ref{Recap_1band}.

The upper critical field is set by the superconducting coherence length, $\Xi$, while the lower critical field is set by the magnetic penetration depth, $\lambda$. Roughly speaking, at \hc1 the whole magnetic flux is contained by a single vortex of radius $\lambda$. A more elaborate treatment take into consideration energy corrections due to the internal structure of the vortex and the Ginzburg parameter $\kappa$,  the ratio of the penetration depth over the superconducting coherence length. When $\kappa$ is large, the ratio $H_{c1} / H_{c2}$ is given with negligible error by the following approximation \cite{tinkham1975,hu1972,klemm1980}:
\begin{equation}
\label{kappa}
\frac{H_{c1}}{H_{c2}} = \frac{\ln (\kappa) + 0.5}{2 \kappa ^2}
\end{equation}

Noting that $H_{c2} = \phi_0 / 2 \pi \xi ^2$, where $\phi_0$ is the quantum of magnetic flux, this leads to the following relation between $H_{c1}$ and $\lambda$ :
\begin{equation}
\label{Hc1}
H_{c1} = \frac{ \phi_0 }{ 4 \pi \lambda ^2 } (\ln (\kappa) + 0.5)
\end{equation}

This widely- used expression is to be used with caution given its implicit assumption of a large $\kappa$. The question is particularly relevant in the case of \STO where the $\kappa$ is relatively low. Harden and Arp \citep{harden1963} made numerical calculations for $\kappa$ values ranging from 0.3 to 100. They found that when $\kappa=5$, $H_{c2}/H_{c1}=22.44$, which is to be compared with t$H_{c2}/H_{c1}=23.74$ deduced from Eq. \ref{kappa}. This represents an error of  6\%. At $\kappa = 10$ the same comparison yields an error of 1.4\%. Therefore, if the ratio $H_{c1}/H_{c2}$ ratio plugged in to Eq. \ref{kappa} yields a $\kappa$ larger than 5, then we can legitimately use Eq. \ref{Hc1} to extract the penetration depth, $\lambda$.

From the measured bulk $H_{c2}$ in three of our samples (combined with unpublished AC susceptibility data in the case of $n_H = 2.6$ \1020 ~sample \citep{lin_unp}), we extracted $\kappa$. When carrier concentration is low and the system is in the clean limit $\kappa\simeq 8.5$. It increases to 17 at high concentration when the mean-free-path shortens and pulls down the coherence length. This allowed us to extract the penetration depth from \hc1 using Eq. \ref{kappa}. Since $\kappa>5$, our use of Eq. \ref{Hc1} is legitimate.

\section{Discussion}
\subsection{Penetration depth}
The superfluid density and the penetration depth are intimately linked through the London equation\cite{tinkham1975}:
\begin{equation}
\label{London}
\lambda^{-2} = \mu _0 e^2 \frac{n_s}{m^{\star}}
\end{equation}

\begin{figure}
\centering
\includegraphics[scale=0.4]{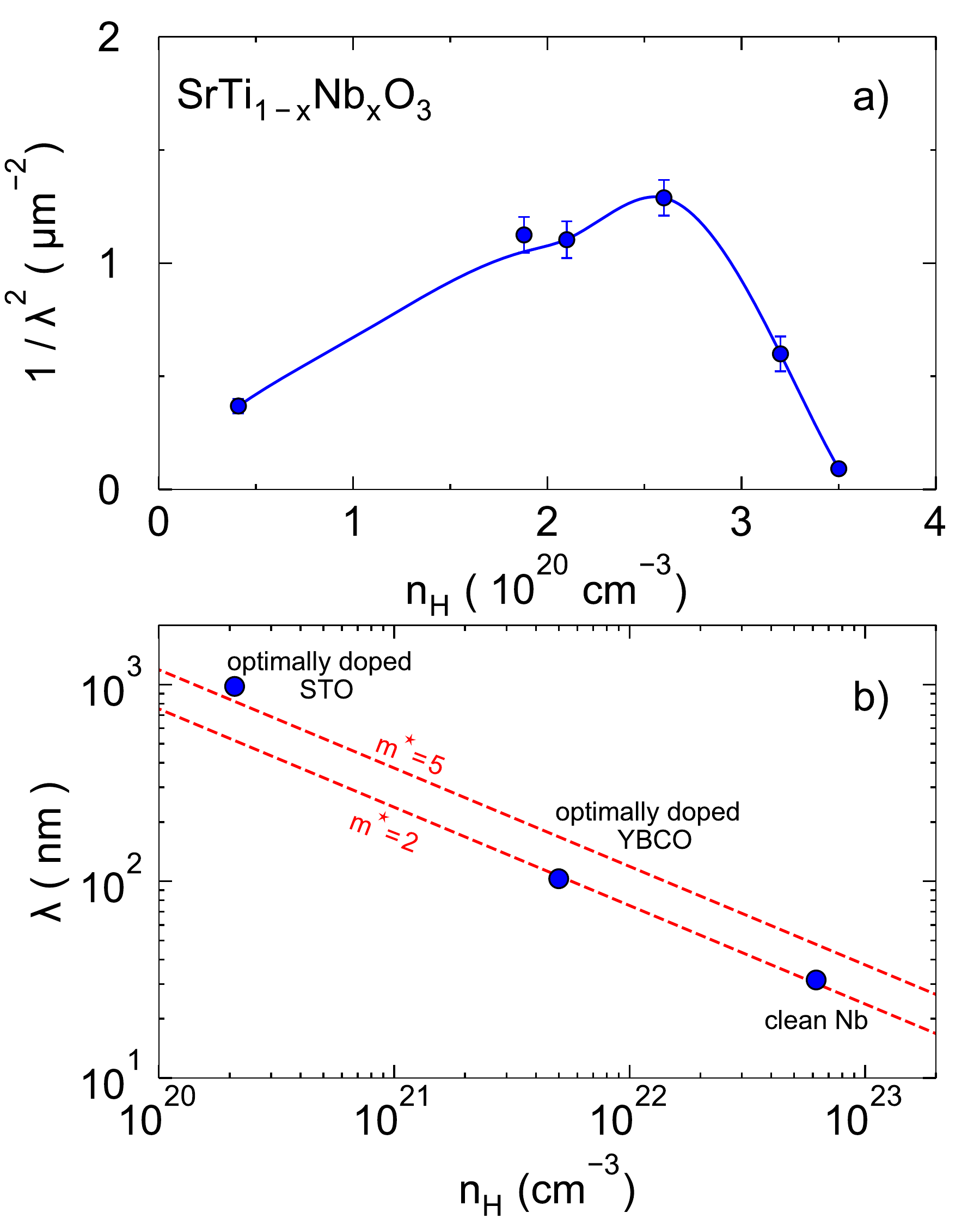}
\caption{\textbf{a}: Doping dependence of the penetration depth plotted as
$\lambda^{-2}$ vs $n_H$. It exhibits a dome-like structure comparable to $T_c$, as in the case of cuprates\cite{lemberger2011}. \textbf{b}: Penetration depth plotted as a function of $n_H / m^{\star}$ in three different superconductors with $m^{\star}$ beeing unitless. The values are detailed in Table \ref{comp_compounds}. The dashed lines represent the expectation of Eq. \ref{London}, assuming $n_H=n_s$. The penetration depth in cuprates is the $ab$ plane one $\lambda_{ab}$}
\label{fig3}
\end{figure}

The doping dependency of $\lambda ^{-2}$ is shown in Fig.\ref{fig3}. It also presents a dome-like structure, reminiscent of the case of cuprates\cite{lemberger2011}. Uemura and co-workers\cite{uemura1989} were the first to notice that the correlation between the superfluid stiffness ($ \propto\frac{n_s}{m^{\star}}$ ) and the critical temperature in cuprates.

\subsection{Comparison with other superconductors}
The penetration depth of optimally-doped strontium titanate is 870nm. Let us now compare this with two other superconductors (Fig.4b). In niobium, it is 31.5nm\cite{varmazis1974}. Given the large difference in their carrier concentration, this is not surprising. The figure compares $\lambda (n_H)$ for these two systems with what is expected according to Eq. \ref{London}, and assuming $n_H=n_s$. One can see that the data points fall close to their expected position. It is instructive to compare the measured $\lambda_{ab}$ in optimally-doped YBCO (103 nm\cite{Pereg2004}) with these two systems. As seen in Fig. 4b, assuming the carrier density given by the Hall coefficient in the vicinity of optimal doping\cite{badoux2016} (See table II for details), this value of the penetration depth is what is expected in this system. This confirms a key expectation of the BCS theory : Absent disorder, what sets the magnitude of the penetration depth in a given superconductor is its carrier density (and not its critical temperature). As far as we know, such an experimental verification of this expectation by a comparison of densities across two orders of magnitude was not done before.

\begin{table}[h!]
\begin{center}
\begin{tabular}{|c|ccccc|}
\hline
\multirow{2}{*}{System} & $T_c$ & $\lambda$ & $m^{\star}$ & $n_S$ & $n_H$ \\
 & (K) & (nm) & ($m_e$) & (10$^{20}$cm$^{-3}$) & (10$^{20}$cm$^{-3}$)\\
\hline
STO  & 0.3 & 952 & 4.2 & 1.3 & 2.1\\
Nb & 9.3 & 31.5\cite{varmazis1974} & 1.5-4\cite{karim1978} & 520  & 620\cite{gilchrist1971} \\
YBCO  & 89 & 103\cite{Pereg2004} & 3.6\cite{ramshaw2015} & 96 & 50\cite{badoux2016}\\
\hline
\end{tabular}
\end{center}
\caption{Penetration depth, superfluid density, normal-state carrier-density in three superconductors. Note that in the case of YBCO, the largest effective mass\cite{ramshaw2015} was measured at p=0.15, below optimal doping (p=0.18) and it is the in-plane penetration depth, $\lambda_{ab}$, which is considered.}
\label{comp_compounds}
\end{table}

\subsection{Disorder }

Disorder can reduce superfluid density through two distinct mechanisms. The first mechanism is relevant to all superconductors. The energy scale of Drude metallic conductivity, $\hbar/\tau$, increases with increasing disorder and decreasing scattering time, $\tau$.  When the superconducting gap, $\Delta$,  is much larger than this energy scale, the condensate density and the density of normal quasi-particles are comparable. In the opposite limit (when $\hbar/\tau \gg \Delta$), a small fraction of normal-state charge carriers have sub-gap energies and one expects $n_{s}\ll n_{H}$\cite{tinkham1959,ferrell1958}.

Quantitatively the superfluid density can be described by the following equation \citep{tinkham1959,ferrell1958}:
\begin{equation}
n_s \simeq \frac{2m^{\star}}{\pi e^2} \int _0 ^{2 \Delta / \hbar} d \omega \sigma_1(\omega)
\label{FGT}
\end{equation}
Where $\sigma_1(\omega)$ is the real part of the optical conductivity.
If we assume that $\sigma_1(\omega)$ behave as a Lorentzian of width $1/\tau$ and zero frequency value $\sigma_0 = ne^2\tau / m^{\star}$ as described by the Drude model we can simply integrate Eq.\ref{FGT}:
\begin{equation}
n_s \simeq \frac{2n}{\pi} \arctan \left( \frac{2\pi \Delta \tau}{\hbar} \right)
\label{FGT_analitique}
\end{equation}
Note that in the dirty limit (\textit{i.e.} $\Delta \tau \rightarrow 0$), leading to the Homes law which states that $n_s/m^{\star} \propto \sigma_0 T_c$ \citep{homes2004,homes2005}.

Another distinct disorder-driven mechanism is specific to nodal superconductors like cuprates and is associated with non-magnetic pair breaking. Note that in both cases, the crucial parameter is ratio of $\hbar/\tau$ to $\Delta$. Since experiments indicate that n-doped SrTiO$_3$ is s-wave\cite{linprb2014,lin2015}, only the first mechanism is relevant here.

\subsection{Superfluid density}
According to band calculations\cite{vdm2011}, metallic strontium titanate has three distinct bands.  This was confirmed by an extensive study of quantum oscillations\cite{lin2014}, finding new frequencies emerging above two critical doping levels. Early tunneling studies\cite{binnig1980} and more recent thermal conductivity measurements\cite{linprb2014} detected two distinct gaps. The cyclotron masses of the three bands are different. The lower band (or the outer sheet of the Fermi surface) is heavier ($m^{\star}_1 = 3.85 \pm 0.35$) compared to the higher bands ($m^{\star}_{2,3} = 1.52 \pm 0.25$). These numbers are consistent with what was found by ARPES ($1.5 m_e$ and $6 m_e$) at a higher concentration\cite{chang2010}. In our window of interest,  three-fourth of all carriers reside in the lower band\cite{lin2014}. The electronic specific heat, $\gamma = 1.55$ mJ mol$^{-1}$ K$^{-2}$ (at $n_H = 2.6$ \1020) is known. Assuming that all carriers reside in one band, one obtains $m^{\star} = 4.2 m_e$\cite{linprb2014}.

\label{1band}

\begin{figure}
\includegraphics[scale=0.4]{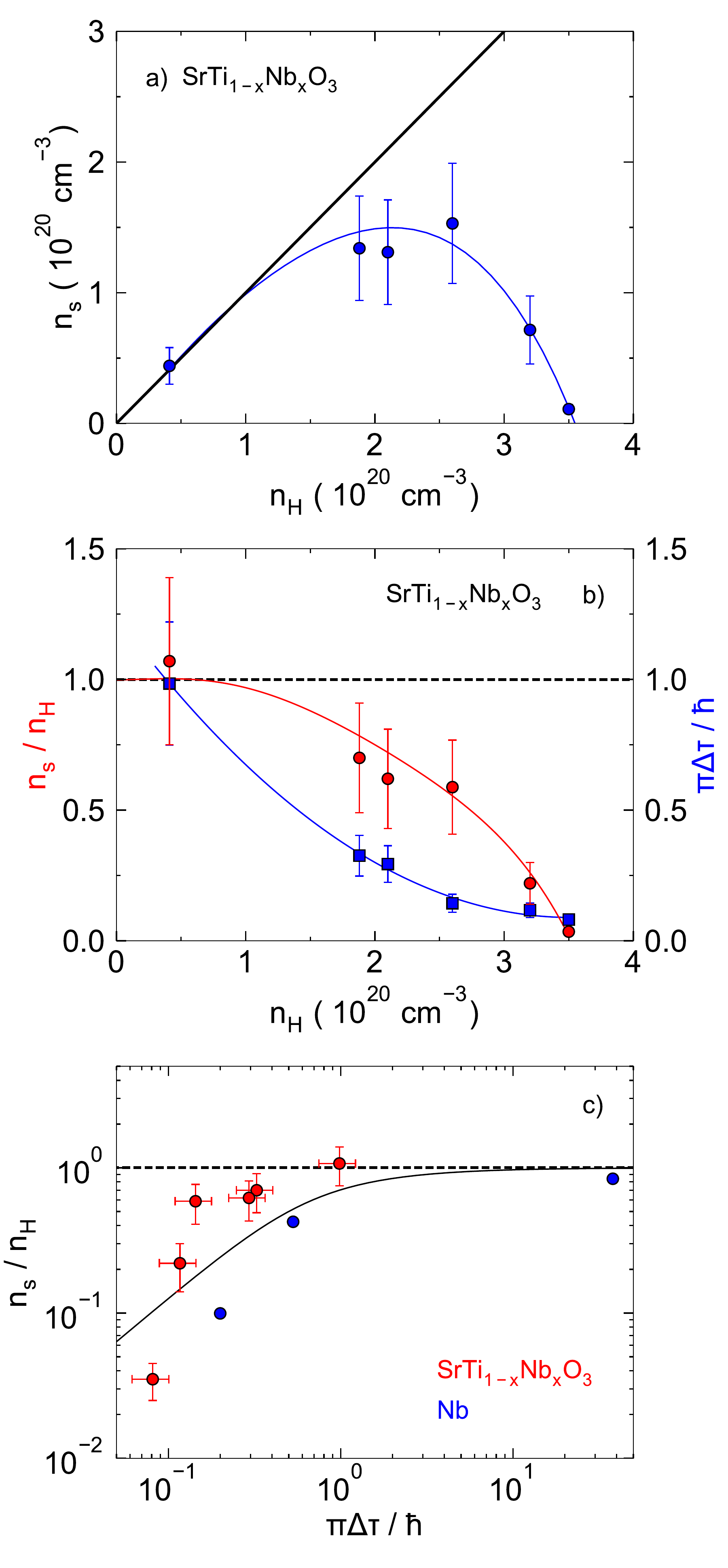}
\caption{\textbf{a}: Evolution of superfluid density as a function of carrier density. The straight line represents $n_S=n_H$. \textbf{b}: The ratio of superfluid density over normal-state carrier density (red discs) compared to the product of the scattering time and the superconducting gap (blue squares) as a function of doping. Lines are guides to the eye. \textbf{c}: The evolution of superfluid density in Nb (blue squares\cite{varmazis1974,klein1994,pronin1998}) and in Nb-doped SrTiO$_3$ (red discs). The continuous black line is the analytic solution of the Ferrell Glover Tinkham sum rule Eq.\ref{FGT_analitique}.}
\label{fig4}
\end{figure}

\begin{table*}
\begin{center}
\begin{tabular}{|c|ccccccccccc|}
\hline
\multirow{2}{*}{Sample} & $n_H$ & $T_c$ & \hc1 & $H_{c2}$ & \multirow{2}{*}{$\kappa$} & $\lambda$ & $n_s$ & $\rho _{2K}$ & $\tau$ & \multirow{2}{*}{$\frac{n_s}{n_H}$} & \multirow{2}{*}{$\frac{\pi \Delta \tau}{\hbar}$} \\
 & (\1020) & (mK) & (Oe) & (Oe) & & (nm) & (\1020) &($\mathrm{\mu \Omega}$.cm) & (ps) & & \\
\hline
&&&&&&&&&&&\\
\multirow{6}{*}{STO} & $0.41$ & $183$~~& $1.6$~~& - & - & $1647$ & $0.44$ & $49$ & $7.43$ & ~$1.07$~ & ~$0.980$~ \\
 & $1.9$ & $300$~~& $4.8$~~& - & - & $943$ & $1.34$ & $53$ & $1.50$ & $0.70$ & $0.326$ \\
  & $2.1$ & $303$~~& $4.7$~~& $240$ & $8.2$ & $952$ & $1.31$ & $71$ & $1.34$ & $0.62$ & $0.294$ \\
  & $2.6$ & $382$~~& $6.05$~~& $480$ & $10.5$ & $881$ & $1.53$ & $109$ & $0.52$ & $0.58$ & $0.144$ \\
  & $3.2$ & $192$~~& $2.8$~~& $210$ & $10.5$ & $1292$ & $0.72$ & $56$ & $0.84$ & $0.22$ & $0.116$ \\
  & $3.5$ & $118$~~& $0.5$~~& $85$ & $17$ & $3305$ & $0.11$ & $45$ & $0.95$ & $0.03$ & $0.081$ \\
\hline
&&&&&&&&&&&\\
\multirow{3}{*}{Nb} & 620 & 9.3 K & 1800 & 4000 & 0.85 & 31.5 & 520 & $2 \, 10^{-4}$ & ~5.8~ & 0.84 & 38 \\
& 620 & 9.3 K & - & - & - & 44 & 264 & 1.2 & 8$\, 10^{-2}$ & 0.42 & 0.53 \\
& 620 & 8.3 K & - & - & - & 90 & 63 & 3.9 & 3$\, 10^{-2}$ & 0.10 & 0.20 \\
\hline
\end{tabular}
\end{center}
\caption{Parameter for \STO, and for clean \cite{varmazis1974,karasik1970} and dirty \cite{klein1994,pronin1998} niobium.}
\label{Recap_1band}
\end{table*}

Let us assume a single band  with an effective mass of 4.2 m$_e$ and  $n_H$ extracted from ( Hall coefficient, which is close to the total number of carriers residing in different Fermi surface pockets and detected by quantum oscillations\cite{lin2014}. The results can be seen in Fig.\ref{fig4}a, which shows the ratio of the extracted $n_s$ over the normal state carrier density $n_H$ for our six samples.  As seen in this figure, the superfluid density and the normal carrier density match each other at low doping. A deviation starts at optimal doping before drastically increasing at higher doping levels. Thus, the superfluid stiffness (that is $\lambda^{-2}$) follows the doping dependence of the critical temperature, because $n_S$ becomes lower than $n_H$ in the overdoped regime.

In order to document the passage to the dirty limit, we extracted the scattering time $\tau$ from low-temperature resistivity ($1/\rho_0 = ne^2 \tau/m^{\star}$) and estimated the magnitude of the superconducting gap using the BCS relation $\Delta \sim 1.76k_BT_c$\cite{binnig1980,thiemann2017}. The product of the two tells us on which side of the clean/dirty limit  the system is. Note that $\ell / \xi = \pi \Delta \tau / \hbar$. When  $\pi \Delta \tau / \hbar <1$, one enters the dirty limit.  Fig.\ref{fig4}b.  compares  the evolution of $n_s/n_H$ and $\pi \Delta \tau / \hbar$.  The  two quantities deviate from unity concomitantly, but are not proportional to each other in the dirty limit as one may expect in the crudest conceivable approximation.  It is also instructive to compare our data with available data on Nb from three different studies\cite{varmazis1974,klein1994,pronin1998}.  Figure \ref{fig4}.c shows the variation of $n_s/n_H$ with $\pi \Delta \tau / \hbar$  in \STO  and in Nb. One can see that in both systems, as theoretically expected\cite{homes2005}, the clean-to-dirty crossover and the loss of superfluid density are concomitant. However, in the case of strontium titanate,  $n_s/n_H$ and $\pi \Delta \tau / \hbar$ are not simply proportional to each other. This is a presumably due to the inadequacy of a single-band approach.

\subsection{Comparison with the interface superconductor}
Let us compare our quantification of superfluid density in the bulk superconducting strontium titanate with the case of the superconducting LaAlO$_{3}$/SrTiO$_{3}$ interface. Bert \emph{et al.}\cite{bert2012} studied how a gate voltage tunes the two-dimensional superfluid density in this system. The maximum superfluid density found was  n$^{2D}_s=3 \times 10^{12}$ cm$^{-2}$. This is much more dilute than the maximum three-dimensional superfluid density found here, which is  n$^{3D}_s= 1 \times 10^{20}$ cm$^{-3}$. The two densities can be contrasted by comparing the average distance between carriers. In their case, the peak n$^{2D}_s$ corresponds to d$_{ee}=5.7nm$. In our case,  n$^{3D}_s$ corresponds to d$_{ee}=2.15nm$. In other words, the peak density for the interface system is significantly more dilute than in the bulk system. The most plausible explanation for this difference is the presence of additional disorder in the interface superconductor. As discussed above, the larger the disorder, the lower the superfluid density.

\section{Summary}

In summary, we found that superconducting \STO has a dome-like  $\lambda^{-2}$ reminiscent of cuprates. Comparing three different systems, one sees that when a superconductor is clean, its $\lambda$ is primarily set by its carrier density and not by its critical temperature.  The density mismatch between superconducting and normal states is concomitant with the entry to the dirty limit. Since in all superconducting domes, the $T_c$ and the gap eventually vanish, the mismatch is expected in any superconducting dome far from optimal doping. Finally, we notice that a quantitative account of disorder-driven loss of superfluid density needs to take into account the multiplicity of the electronic bands and the superconducting gaps.

\section{Acknowledgments}

We thank A. J. Millis, J. Orenstein and L. Taillefer for helpful discussion and comments. This work is supported by Agence Nationale de la Recherche through the QUANTUM LIMIT project, by Fonds-ESPCI-Paris  and by JEIP-Coll\`{e}ge de France.

\bibliography{bib}
\end{document}